\documentclass{vgtc}                          

\ifpdf
  \pdfoutput=1\relax                   
  \usepackage{graphicx}                
  \DeclareGraphicsExtensions{.pdf,.png,.jpg,.jpeg} 
\else
  \usepackage{graphicx}                
  \DeclareGraphicsExtensions{.eps}     
\fi%

\graphicspath{{figures/}{pictures/}{images/}{./}} 

\usepackage{microtype}                 
\PassOptionsToPackage{warn}{textcomp}  
\usepackage{textcomp}                  
\usepackage{mathptmx}                  
\usepackage{times}                     
\usepackage{cite}                      
\usepackage{tabu}                      
\usepackage{booktabs}                  

\usepackage{xcolor}
\usepackage{lipsum}
\usepackage{hyperref}
\usepackage{multirow}

\newcommand\XX[1]{{\color{black}{#1}}}

\onlineid{1109}

\vgtccategory{Research}

\vgtcinsertpkg

\title{The Anatomical Edutainer}

\author{Marwin Schindler\\ %
        \scriptsize TU Wien, Austria %
\and Hsiang-Yun Wu\\ %
     \scriptsize TU Wien, Austria %
\and Renata G. Raidou \thanks{mail@renataraidou.com \vspace{-40pt}}\\ %
     \parbox{2in}{\scriptsize \centering University of Groningen, The Netherlands \\ TU Wien, Austria}}

\teaser{
\centering{
\begin{tabular}{ccc}
\includegraphics[height = 1.8in]{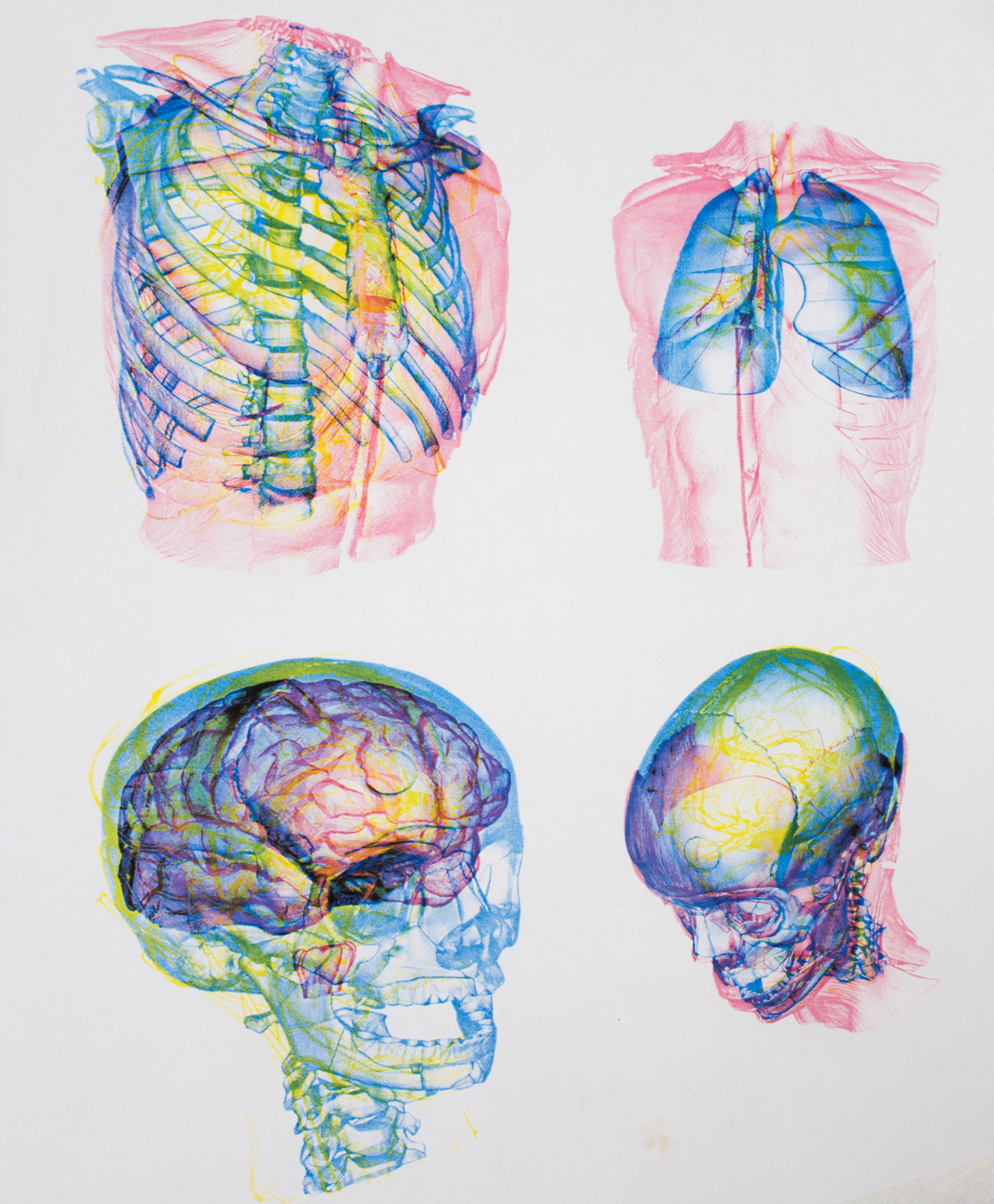} &
\includegraphics[height = 1.8in]{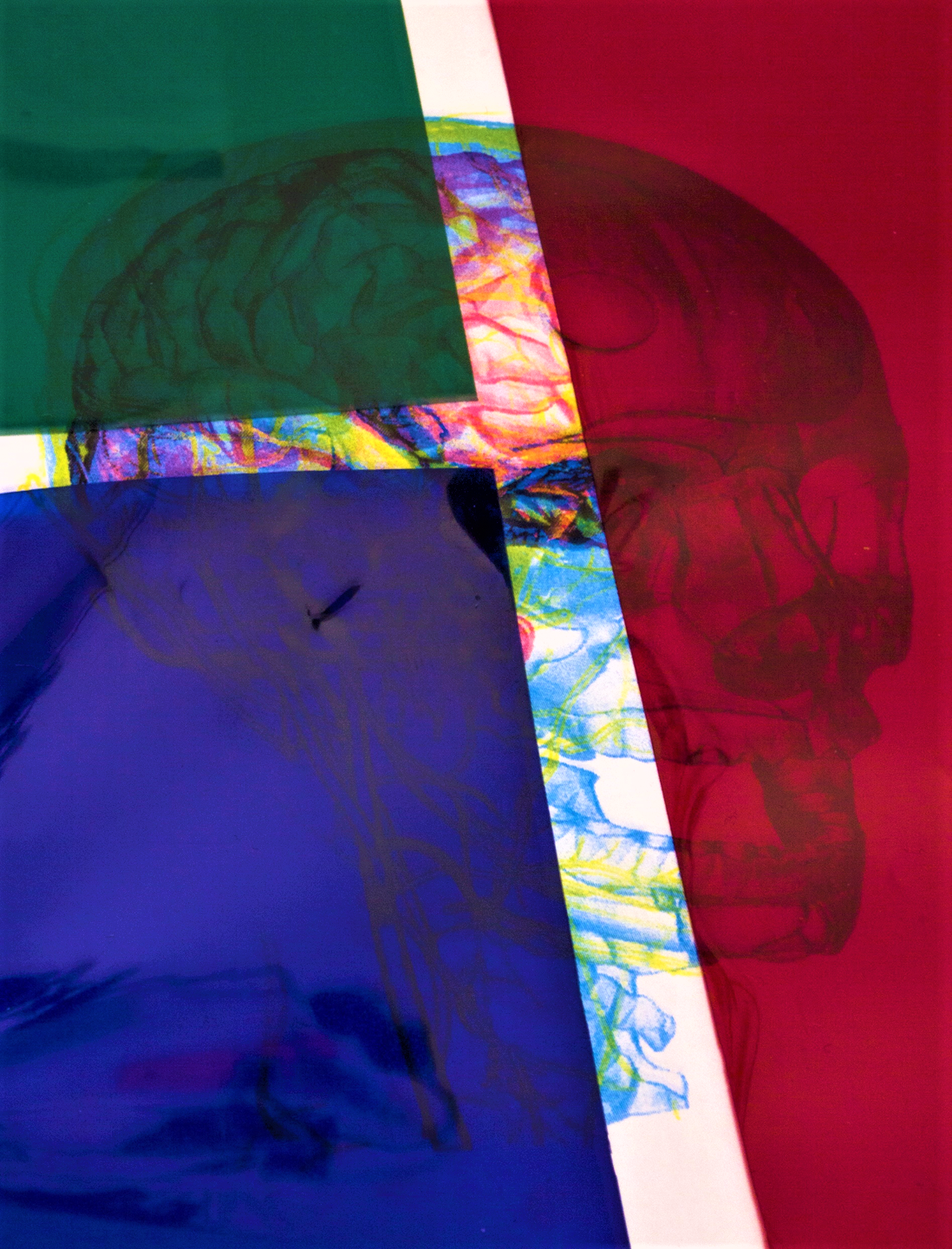} &
\includegraphics[height = 1.8in]{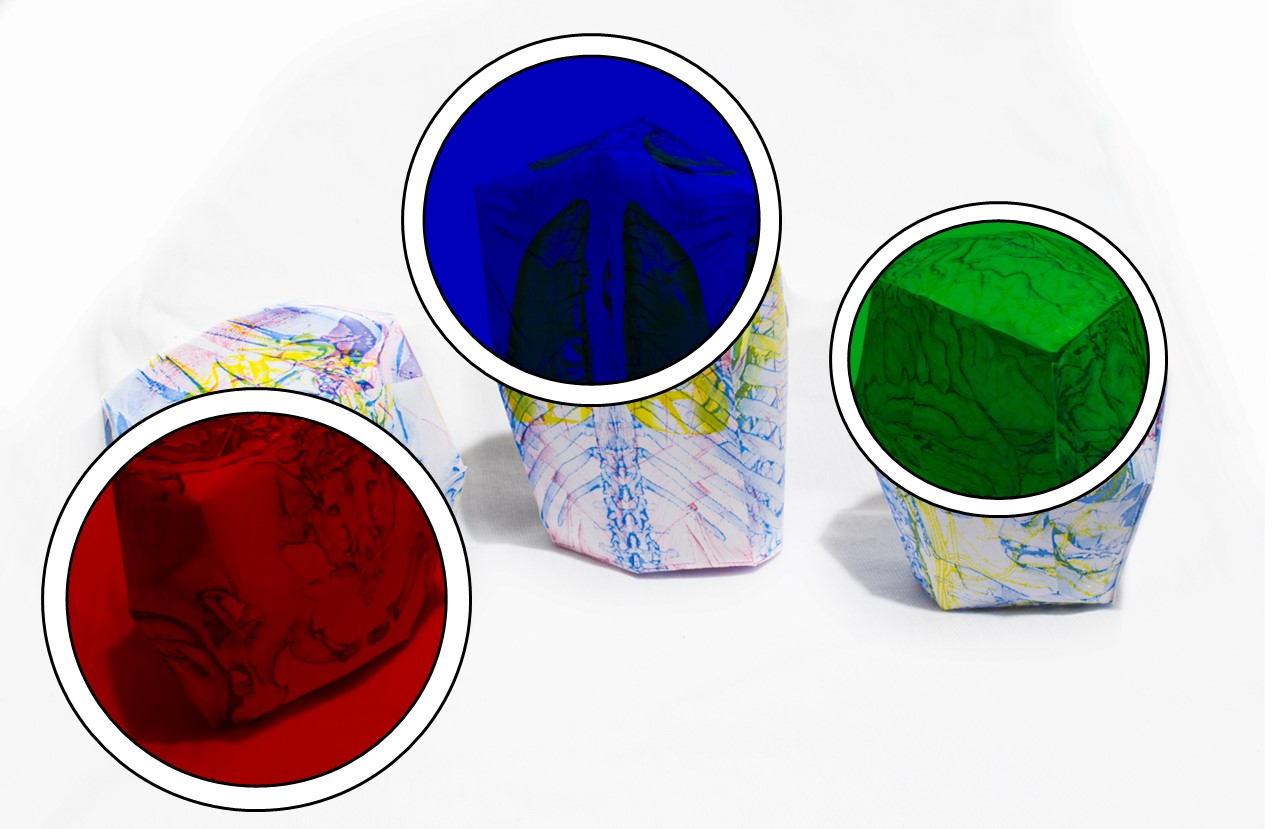} \\
(a) & (b) & (c) \\
\end{tabular}
\caption{Examples generated with the \emph{Anatomical Edutainer}. (a) Printable 2D physicalizations of the human torso and head. In each representation, individual anatomical structures are matched to individual hues, which can be isolated (b) with the use of colored filters. (c) Foldable 3D physicalizations of the human pelvis, torso and head. Similarly to the 2D case, distinct hues are assigned to individual anatomical structures, which can be isolated with the use of colored lights (or colored filters, as in (b)).}
\label{fig:teaser}
}}

\abstract{
Physical visualizations \XX{(}i.e., data representations by means of physical objects\XX{)} have been used for many centuries in medical and anatomical education. 
Recently, 3D printing techniques started also to emerge. 
Still, other medical physicalizations that rely on affordable and easy-to-find materials are limited, while smart strategies that take advantage of the optical properties of our physical world have not been thoroughly investigated. 
We propose the \textit{Anatomical Edutainer}, a workflow to guide the easy, accessible, and affordable generation of physicalizations for tangible, interactive anatomical edutainment. 
The \textit{Anatomical Edutainer} supports 2D printable and 3D foldable physicalizations that change their visual properties \XX{(}i.e., hues of the visible spectrum\XX{)} under colored lenses or colored lights, to reveal distinct anatomical structures through user interaction.
} 

\CCScatlist{
  \CCScatTwelve{Human-centered computing}{Visu\-al\-iza\-tion}{Visu\-al\-iza\-tion application domains}{Scientific visualization};
  \CCScatTwelve{Applied computing}{Life and medical sciences}{}{}
}

\begin{document}


\firstsection{Introduction} \label{sec:intro}
\maketitle

Visualizations function as a major instrument for exploring data and for obtaining insights into relevant research questions, while they can also be useful within an \emph{educational} or \emph{edutainment} context.
\textit{Physical visualization}~\cite{PhysicalVisualization} is a subdomain of classical visualization, where the data is represented and explored by means of \emph{physical objects}, instead of being displayed on-screen. 
These physical objects (or \textit{physicalizations}) are tangible \XX{and} employ additional senses, such as touch, to enhance visual cognition and perception. 
Physicalizations have been proven to provide a high degree of engagement, enabling the user to focus on the physical representation of the underlying data and, therefore, \XX{to understand and remember} the embedded information better~\cite{jansen2013evaluating,SS15}. 
This property can be further supported by offering additional interactivity and \XX{amusement}---for instance, through shapeable objects~\cite{S15}. 

Among other applications, physical visualizations have also been used  within the field of \textit{medical education}. 
Examples of medical physicalizations include model kits that showcase the human anatomy, or contextual 3D printed models~\cite{ANG201942}. 
Both examples \XX{may be} suitable for training clinical personnel, but not so suitable for non-experts, as their manufacturing process is rather costly and/or complex. 
Medical physicalizations that rely on affordable materials and that are available to everyone, are still limited.  
Additionally, smart strategies that make better use of the optical properties of our physical world have not been investigated in this context.

The contribution of this work is the design of a workflow for the easy, accessible and affordable generation of physicalizations that can be used within the context of tangible, interactive anatomical edutainment of the general population \XX{(}e.g., schoolchildren\XX{)}. 
As opposed to 3D printing approaches, our resulting physicalizations make use of affordable materials that are widely available, such as paper, colored foils, and light. 
Our workflow, which we named \textit{Anatomical Edutainer}, consists of two main components: 
\vspace{-5pt}

\begin{itemize}
\item{The generation of \textit{2D printable physicalizations}, which exhibit different visual properties \XX{(}i.e., hues of the visible spectrum\XX{)} under colored lenses or colored lights, and reveal distinct anatomical structures.}\vspace{-5pt}
\item{The generation of \textit{3D foldable physicalizations}, where anatomical structures undergo an unfolding step, to ensure that they can be printed and assembled to a 3D papercraft. The assembled papercraft can be \XX{subsequently} explored under colored lenses or lights, similarly to its 2D counterpart.}
\end{itemize}
\vspace{-5pt}

Anyone with access to a computer and a common printer is able to create our proposed physicalizations, while the colored filters or lights are widely available and affordable. 
Additionally, \XX{the templates of} our physicalizations need to be created only once and can be easily reprinted, which makes them an affordable and accessible tool for educational purposes\XX{, such as at art exhibitions or science museums.} 
The tangible character of the \XX{3D papercraft assembly} adds to the enjoyment \XX{of the process}, making them especially \XX{suitable for} children anatomical edutainment.
\begin{figure*}[!t]
\begin{minipage}{0.45\textwidth}
\centering{
 \setlength{\tabcolsep}{10pt}
\begin{tabular}{ccc}
\multicolumn{2}{c}{\includegraphics[height = 0.27\textwidth]{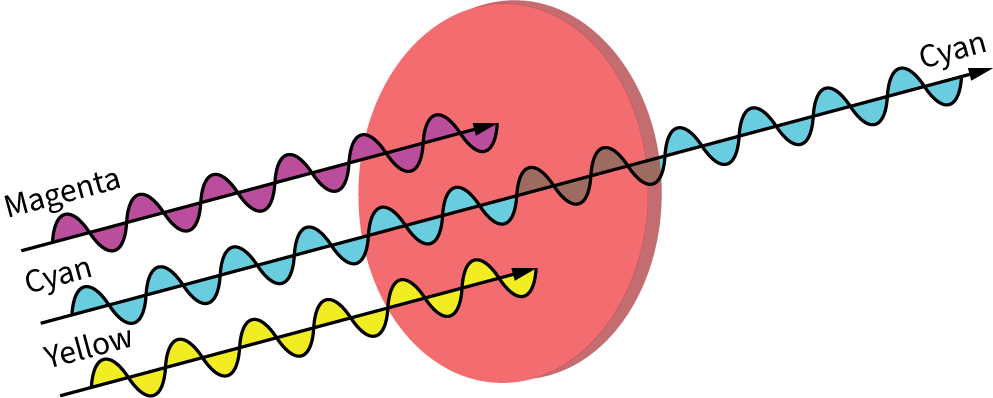}}
\vspace{-3pt} \\
\multicolumn{2}{c}{(a)}\\
\includegraphics[height = 0.35\textwidth]{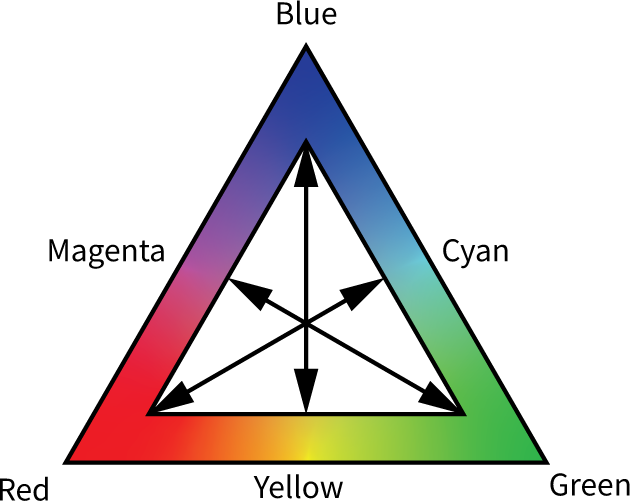} &
\includegraphics[height = 0.35\textwidth]{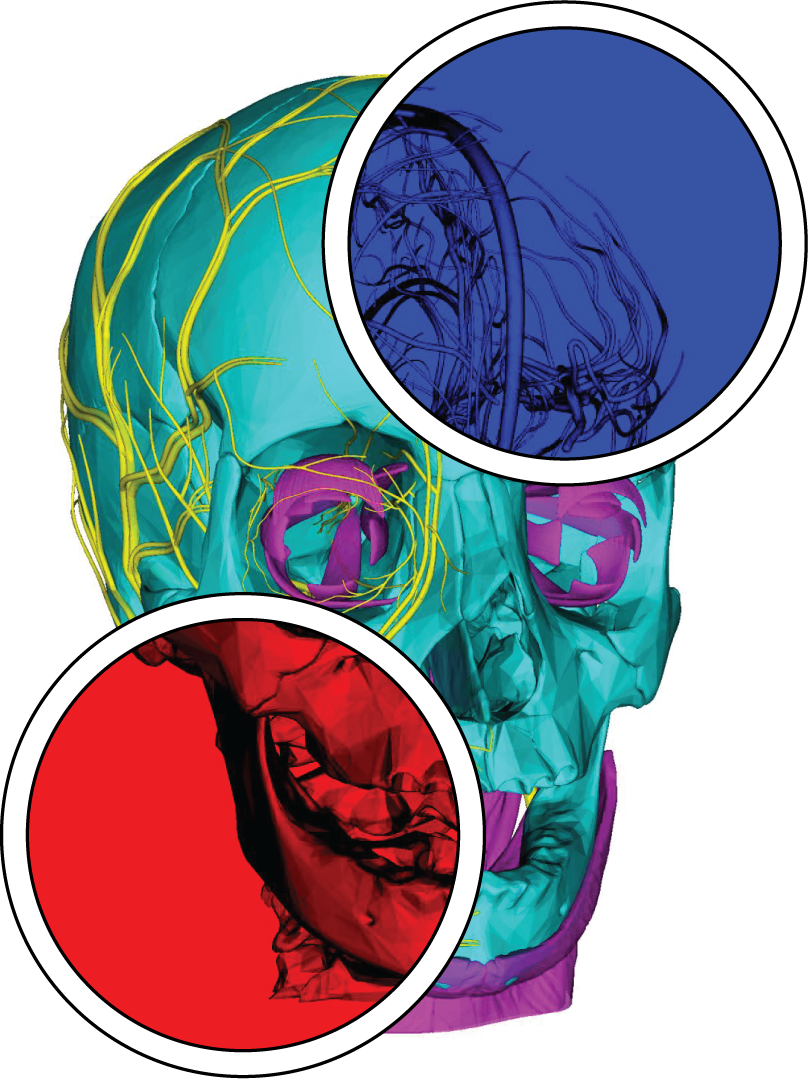}
\vspace{-1pt} \\
(b) & (c) \\
\end{tabular}
}
\caption{Basic concepts \XX{of} the \textit{Anatomical Edutainer}. (a) Example of a red color filter, preserving only the cyan light. (b) The color compensation triangle, showing pairs of colors that interact. (c) A red and a blue color filter applied on a rendering, to unveil the skull (in cyan) and the blood vessels (in yellow), respectively.}
\label{fig:concepts}
\end{minipage} %
\begin{minipage}{0.01\textwidth}
\parbox{0.01\textwidth}{}
\end{minipage} %
\begin{minipage}{0.53\textwidth}
	\centering
	\includegraphics[width=1.0\linewidth]{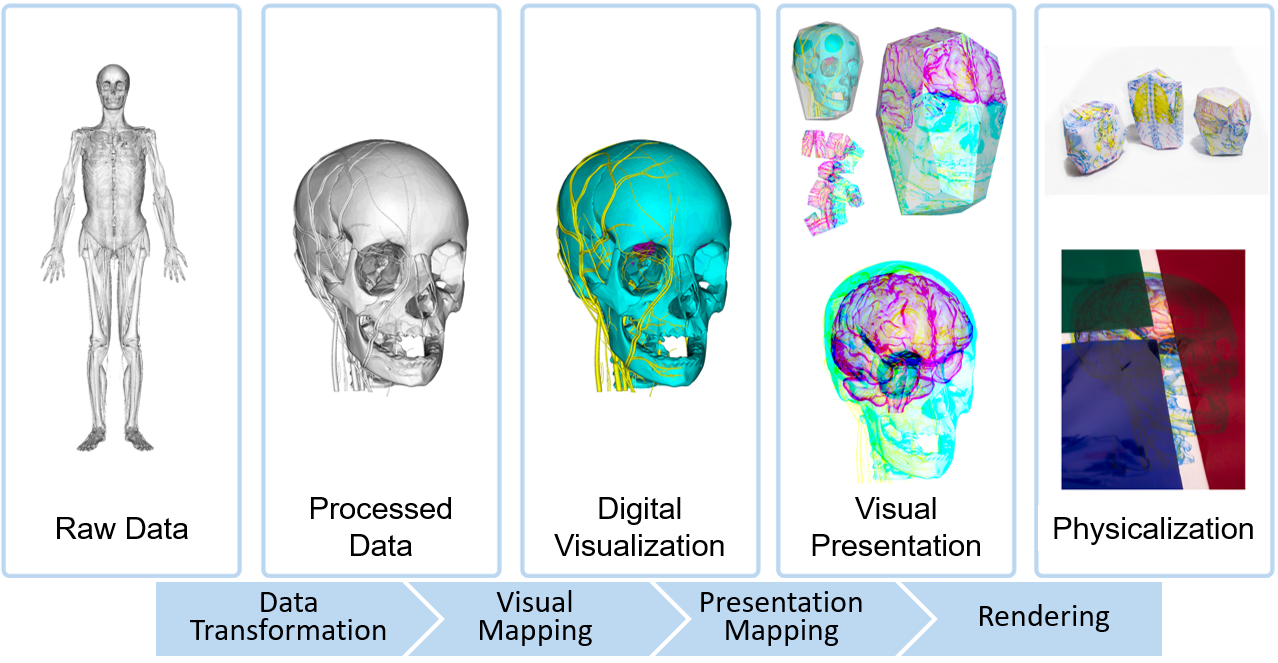}
	\caption{The workflow of the \textit{Anatomical Edutainer}. \emph{Data Transformation} processes the raw meshes. \emph{Visual Mapping} produces the digital visualizations and \emph{Presentation Mapping} prepares the visual presentations according to the colored filters. \emph{Rendering} generates the 2D and 3D physicalizations.}
	\label{fig:1}
\end{minipage}
\vspace{-10pt}
\end{figure*}

\section{Related Work}
\label{sec:related}

Recently, Preim and Saalfeld presented a survey on virtual anatomy education systems~\cite{preim2018survey}, focusing mainly on systems for medical students.
Although we focus on applications for the general population, there is a wide range of classical on-screen visualization techniques for anatomical education and training, using surface and volume rendering~\cite{pommert2001creating, vazquez2008interactive, halle2017open, smit2016online}, as well as virtual and augmented reality~\cite{saalfeld2016semi, pohlandt2019supporting}.
When looking into the theories of \textit{constructivism}~\cite{huang2010investigating} and \textit{embodiment}~\cite{jang2017direct}, it becomes evident why the investigation of data physicalization approaches that go beyond the screen space suits anatomical edutainment.
According to constructivism, active learning supports knowledge construction with less cognitive load, while embodiment supports that learning can benefit from the involvement of physical interaction.
Recently, experiments in the domain of physicalization have been conducted to demonstrate the benefits of mapping data to objects and their physical world properties~\cite{SS15}.
\XX{Physicalization has even been applied to types of data which are not inherently 3D, such as typical barcharts~\cite{add1,add3,add4,jansen2013interaction}, to stimulate the engagement of the intended users.}
Physical representations may either aid or replace digital representations, while the underlying data can be explored with additional senses---going beyond the optical channel~\cite{moere2008beyond,jansen2015opportunities}.
In certain cases, physical visualizations have proven to be even more effective than their virtual counterparts~\cite{moere2008beyond, jansen2013evaluating}.

The last few years, medical data physicalization has been mainly driven by 3D printing techniques~\cite{mcmenamin2014production, Hybrid3Dprinting, 3Dprintingbasedonimagingdata, CardiacBloodFlowPhys}.
Recent 3D printing advancements allow nowadays more and more complex digital fabrications of anatomical models\XX{---for instance,} the creation of a tangible physicalization of cardiac blood flow~\cite{CardiacBloodFlowPhys}.
Despite its high educational value, 3D printing remains time-consuming, complex, and quite expensive.
Medical data physicalizations that do not involve 3D printing, and use other materials are still limited~\cite{markovicDevelopment, olry2000wax}.
A recent cost-effective and approachable paper-based physicalization includes the generation of volvelles~\cite{Vol2velle}.
These are interactive wheel charts of concentric, rotating disks that mimic on-screen volume rendering for non-expert users.
Finally, an important source of inspiration for this work has been the book \textit{Illumanatomy}~\cite{Carnovsky}.
Here, red, green and blue filters are used on printed visualizations (colored in cyan, magenta and yellow), to allow a different gamut of light to pass and to mask the remaining wavelength range. 
\section{The Anatomical Edutainer}
\label{sec:method}

\subsection{Basic Concepts}
\label{ssec:concepts}
The \textit{aim} of this work is the design of a workflow for the generation of tangible physical visualizations of the human anatomy, which take advantage of the optical properties of our physical world, and combine entertainment with education.
A basic \textit{requirement} in our approach is the accessibility of the resulting physical visualizations, as they should be easily producible, while keeping the necessary materials \XX{and} their costs, as low as possible.
The \textit{fundamental concept} of these physical visualizations is the ability to inspect the generated physical models under appropriate colored lenses (or colored lights) to unveil distinct (and potentially nested) anatomical structures.
To this end, we exploit the properties of light and the light-modulating properties of a typical color lens, as shown in Figure~\ref{fig:concepts}(a).
Since we investigate printable physical visualizations, we employ the subtractive color printing properties~\cite{berns2019billmeyer}.
This means that cyan, magenta, and yellow inks (subtractive) control respectively the red, green, and blue channels (additive), as reflected from a white paper.
When two subtractive primaries overlap, an additive primary is produced \XX{(}e.g., cyan and magenta produce blue\XX{)}.
An overlap of all subtractive inks will appear black.
The color combinations that interact most efficiently are illustrated in Figure~\ref{fig:concepts}(b).

\XX{Colored lenses can be employed to isolate separate anatomical structures~\cite{Carnovsky}, if those have been assigned to distinctly colored ``layers'',} as shown in the example of Figure~\ref{fig:concepts}(c).
Here, the bones, which are colored in cyan, can be seen under a red filter, while all other structures are hidden.
Similarly, under a blue filter, only the yellow colored blood vessels are preserved.
With this approach, a colored lens is required for interacting with the physical visualizations.
Colored lenses are widely used in photography, but they can be expensive.
Hence, simple colored transparent foils or colored lights that are easier to buy and more affordable, are sufficient.

\subsection{Methodological Workflow}
\label{ssec:method}
The steps of our workflow are conceptually depicted in Figure~\ref{fig:1}.
In the first step (\textit{\textbf{Data Transformation}}), we convert the raw data to closed triangular meshes, before importing them into our system.
These are generic anatomical 3D models, representing different parts of the human body, which we acquired from the BodyParts3D database~\cite{mitsuhashi2009bodyparts3d}.
If personalized segmentations from medical imaging data (e.g., CT or MRI) would be available, then these could be employed instead.
This first step also involves filtering \XX{(}i.e., the selection of the relevant meshes for the visualization\XX{)}, and potential resampling, as the meshes from BodyParts3D often consist of $>1M$ triangles for large structures, such as the skull.

In the second step (\textit{\textbf{Visual Mapping}}), the user can adjust the visual properties of the meshes, to optimize their visual attributes based on the available colored filters.
In this work, a combination of cyan, magenta and yellow can only be used for the meshes.
Other colors \XX{would} not give a sufficient differentiation under primary \XX{(}i.e., red, green and blue\XX{)} colored lenses or lights, which are easier to obtain.
An alternative would be to \XX{assign} the red, green and blue channels \XX{to} the meshes, but their blending would require a black background, which is not suitable for printing.
\XX{Although in all our examples we use a consistent color mapping (cyan for bones, magenta for soft tissues or organs, and yellow for the cardiovascular system), this choice is neither exclusive, nor important.
Each structure can only be isolated and observed under the respective colored filter.
The filter will show the preserved structure as black, it will cut off the other two structures, and show the background in the filter's same color, as shown in Figure~\ref{fig:concepts}(c).}
In the future, we plan to investigate the use of other advanced color combinations.
For the generation of a 2D physical visualization, camera adjustments are also required for an optimal---according to the user---view on the anatomical structures.

\begin{figure*}
\centering{
\setlength{\tabcolsep}{1pt}
\begin{tabular}{ccccc}
\includegraphics[height = 1.4in]{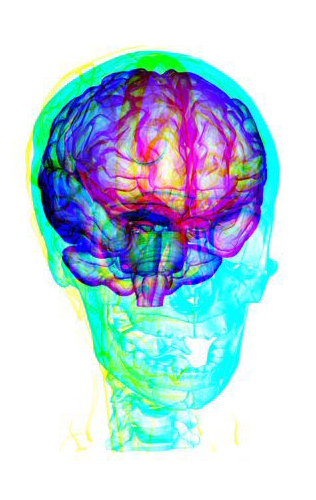} &
\includegraphics[height = 1.4in]{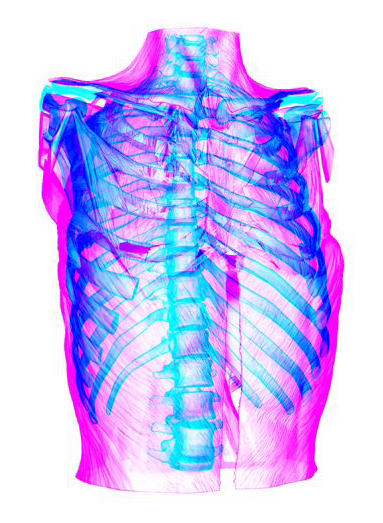} &
\includegraphics[height = 1.4in]{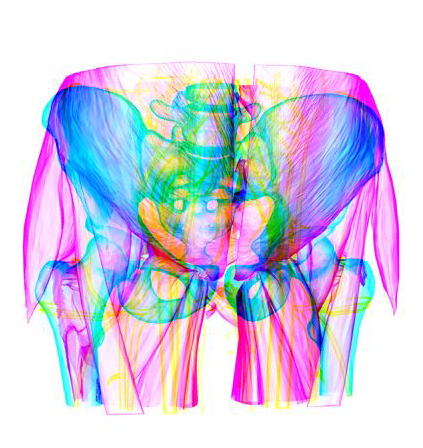} &
\includegraphics[height = 1.4in]{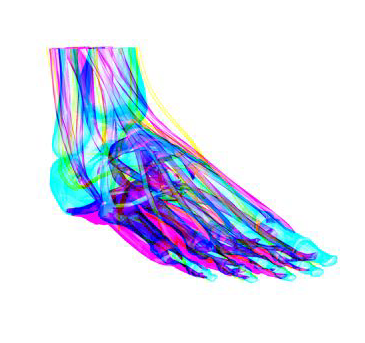} &
\includegraphics[height = 1.4in]{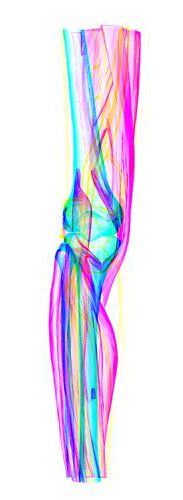} \\
(a) & (b) & (c) & (d) & (e) \\
\includegraphics[height = 1.4in]{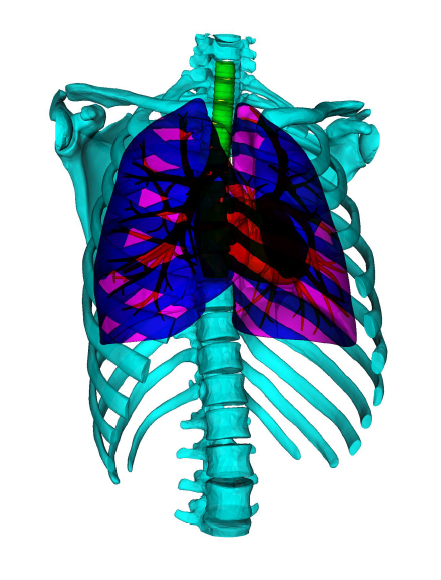} &
\includegraphics[height = 1.4in]{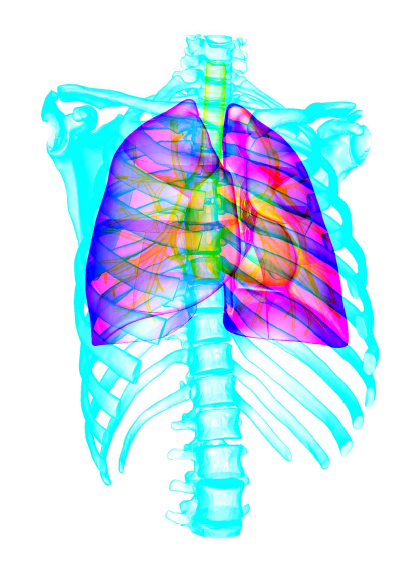} &
\multicolumn{3}{c}{\includegraphics[height = 1.4in]{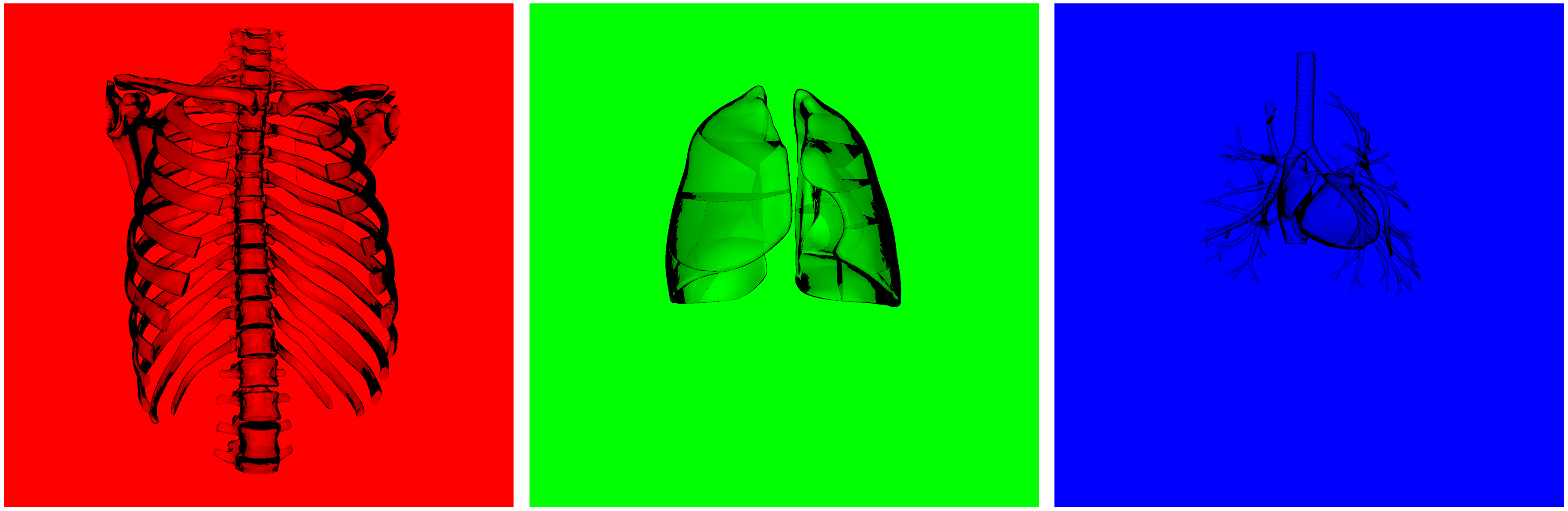}} \\
(f) & (g) & \multicolumn{3}{c}{(h)}
\end{tabular}
}
\caption{Printable 2D physicalizations of (a) a head, (b) a torso, (c) a pelvis, (d) a foot and (e) a leg. All physicalizations comprise three distinct structures, encoded in cyan, magenta or yellow. In (f)-(g), we show the effect of the depth peeling and brightening conducted during the \textit{Presentation Mapping} for a torso. In (h), we show the application of three digital filters for isolating the underlying structures.}
\label{fig:result2d}
\end{figure*}

In the next step (\textit{\textbf{Presentation Mapping}}), we aim to preserve all nested and hidden structures in the rendered view.
First, depth peeling is employed to ensure that each mesh is rendered in the correct order.
This enables an accurate rendering of the meshes without sorting the triangles, by peeling the geometry from front to back.
This is important for combining correctly the images of the individual meshes to form a single image---even for more complex cases with intersecting semi-transparent objects.
Then, color blending is employed, for the color multiplication of all overlapping structures, according to the subtractive scheme, discussed before.
Examples thereof are presented in Figure~\ref{fig:result2d}(a-e).
As the renderings tend to get dark with multiple overlapping color values---especially, due to the use of over-saturated colors---normalization and brightening is performed, as depicted in Figure~\ref{fig:result2d}(f-g).

When the users are pleased with the current visual representation \XX{(}i.e., the chosen meshes colors and camera settings\XX{)}, they can move on to the next step (\textit{\textbf{Rendering}}).
Here, the \textit{Anatomical Edutainer} workflow branches into two different physical visualization techniques---\textit{\textbf{2D}} and \textit{\textbf{3D}}.
\XX{The choice is left to the user, to either print the previously generated 2D visual representation, or to create a 3D papercraft.}
The 2D physical visualizations offer some degree of interactivity, when explored under the colored lenses.
Yet, a higher degree of engagement, understanding and entertainment could be offered with a 3D physical model~\cite{S15,Korpitsch2020}.
Also, 3D physical visualizations preserve better the 3D spatial context of anatomy.

\begin{figure}[t]
\centering{
 \setlength{\tabcolsep}{1pt}
 \begin{tabular}{ccc}
\multicolumn{3}{c}{\includegraphics[height = 1.3in]{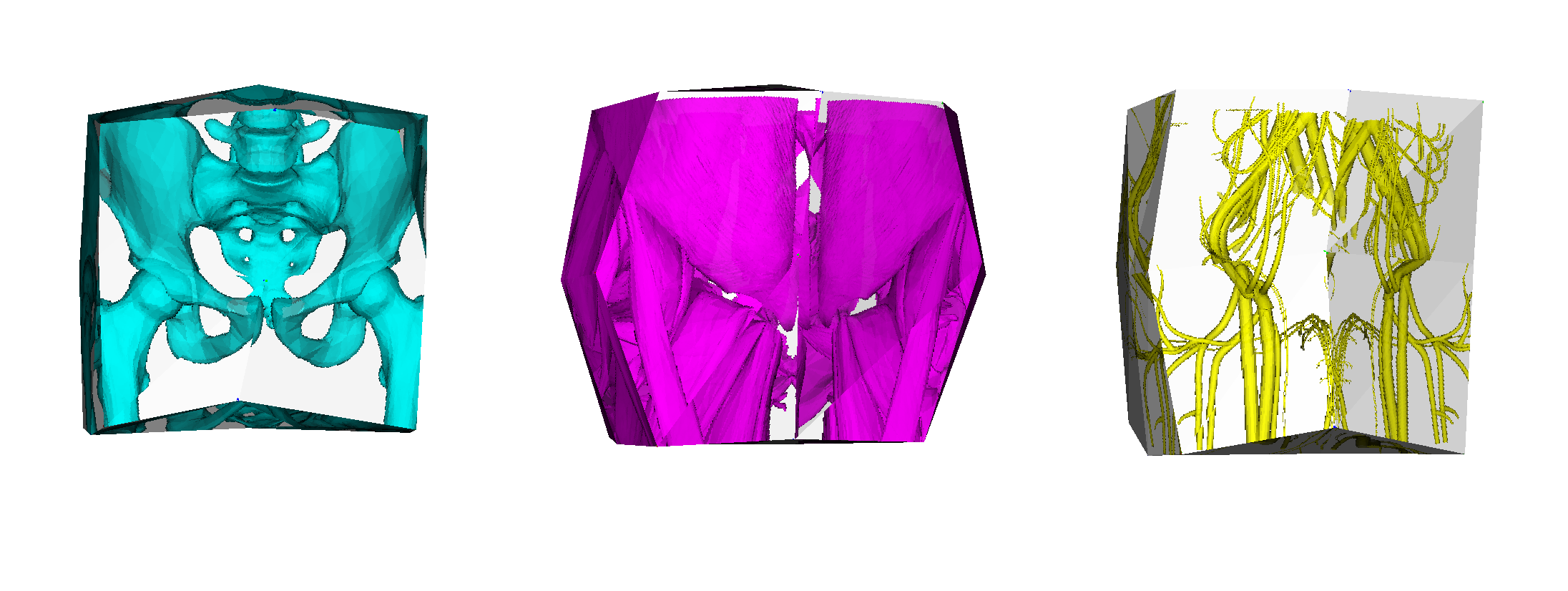}}
\vspace{-16pt} \\
\multicolumn{3}{c}{(a)} \\
\includegraphics[height = 1.2in]{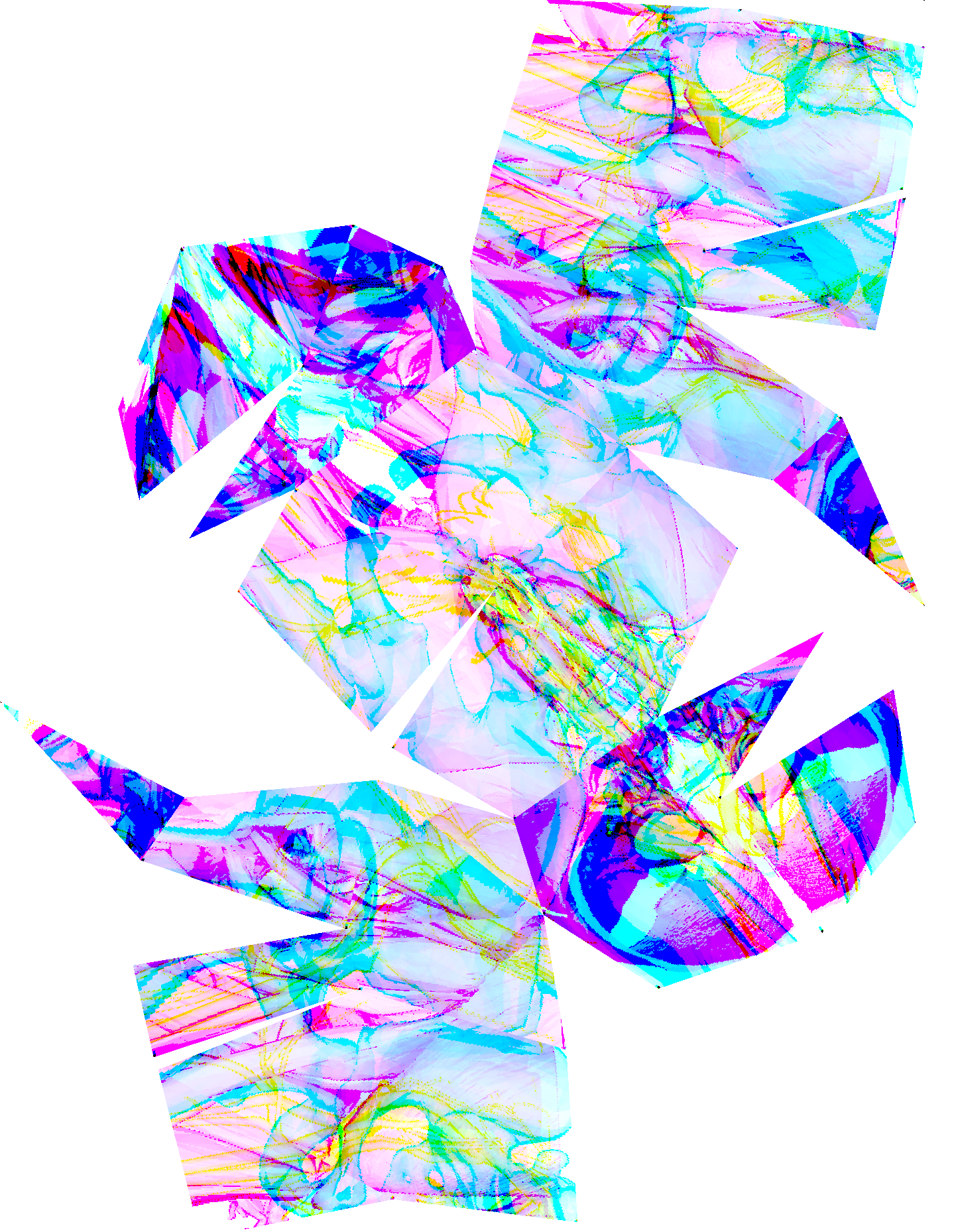} &
\includegraphics[height = 1.2in]{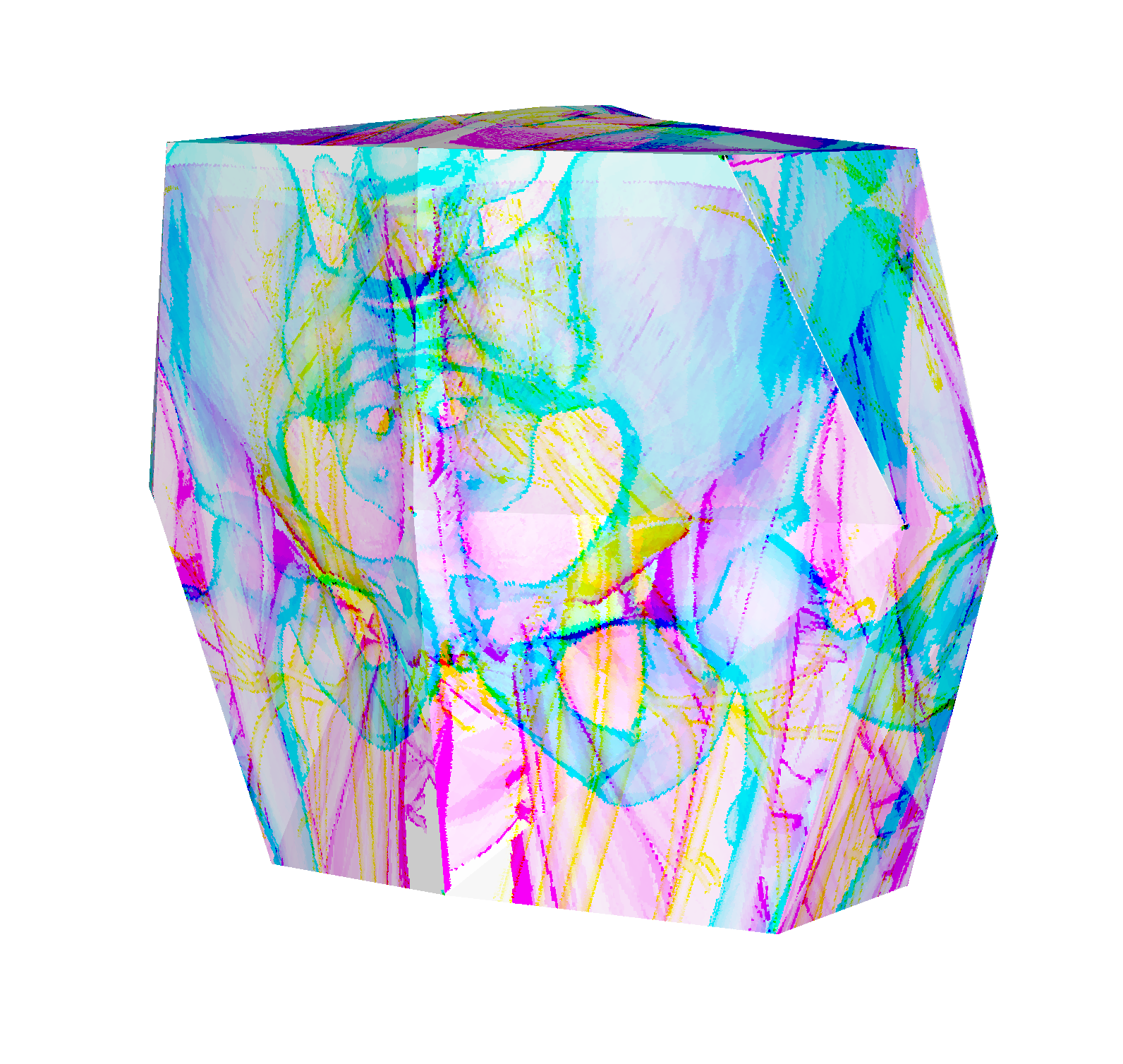} &
\includegraphics[height = 1.2in]{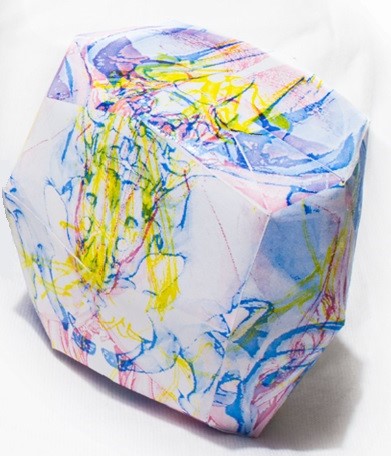} \\
(b) & (c) & (d)
\end{tabular}
}
\caption{Foldable 3D physicalization of a pelvis, where the bones are encoded in cyan, the muscles in magenta and the blood vessels in yellow. A common paper mesh template is generated, then wrapped around each individual structure (a), and used to create a combined texture (b). This is applied back to the combined paper mesh (c) that can be printed and folded into a 3D papercraft (d). \vspace{-8pt}}
\label{fig:result3da}
\end{figure}

For the 3D papercraft generation, a bounding box enclosing all structures is extracted, triangulated, and subdivided.
We move each of the vertices of the subdivided bounding box towards the closest surface point among all anatomical structures, to create a coarse and approximate surface mesh that encloses all of them.
This is the basis for the papercraft and we refer to it, as the ``paper mesh''.
This paper mesh is then unfolded to a planar mesh, and the coordinates of the planar mesh are assigned as UV coordinates to the paper mesh.
Then, we generate the texture of the paper mesh.
The paper mesh is first duplicated and wrapped around each individual anatomical structure (Figure~\ref{fig:result3da}(a)).
We refer to these as the ``wrapped paper meshes''.
For each triangle in each wrapped paper mesh, we render a frame and we concatenate it onto a dedicated texture image.
Using the UV coordinates of the paper mesh, we map the triangles of the previously unfolded planar mesh to the corresponding triangles of the wrapped paper meshes to generate a texture for each anatomical structure.
The resulting textures of the structures are subsequently combined (Figure~\ref{fig:result3da}(b)) in the same way as in the 2D case, and the unfolding is digitally reassembled (Figure~\ref{fig:result3da}(c)).
Triangles with perspective-related artifacts can be locally selected in the application interface (if needed) and flattened to a mean plane \XX{(}i.e., a plane with an origin at the center of the selected triangle points, and a normal that is equivalent to the average normal of all selected points\XX{)}, to reduce discontinuities.
The unfolded mesh is afterwards printed, cut-out and folded back to a 3D papercraft (Figure~\ref{fig:result3da}(d)).

\XX{The \textit{Anatomical Edutainer} is implemented in Python 3.8 with the use of \hyperref[www.vtk.org]{VTK} and PyQt5, while the mesh unfolding (from paper mesh to planar mesh) employs \hyperref[www.blender.org]{Blender}.}

\section{\XX{Results and Discussion}} \label{sec:result}
Figure~\ref{fig:teaser} \XX{depicts} examples of physicalizations in 2D and in 3D, as resulting from the \textit{Anatomical Edutainer} workflow.
In Figure~\ref{fig:teaser}(a), we demonstrate how 2D physicalizations from different parts of the human body \XX{(}i.e., torso and head\XX{)} look like after printing.
These 2D physicalizations support the exploration of up to three different anatomical meshes, with the use of three different hues \XX{(}i.e., magenta, cyan and yellow\XX{)}.
In Figure~\ref{fig:teaser}(b), we show how three filters \XX{(}red, green and blue\XX{)} are used to selectively inspect different structures. 
As discussed in Section~\ref{ssec:concepts}, the blue filter preserves the structures rendered in yellow, the red filter preserves cyan structures and the green filter the magenta ones. 
For the foldable 3D physicalizations in Figure~\ref{fig:teaser}(c), a similar interaction and exploration approach can be used \XX{(}i.e., based on preferred color filters or colored lights\XX{)}.

\begin{figure}[t]
\centering{
 \setlength{\tabcolsep}{1pt}
 \begin{tabular}{cccc}
\includegraphics[height = 1.2in, width=0.15\textwidth]{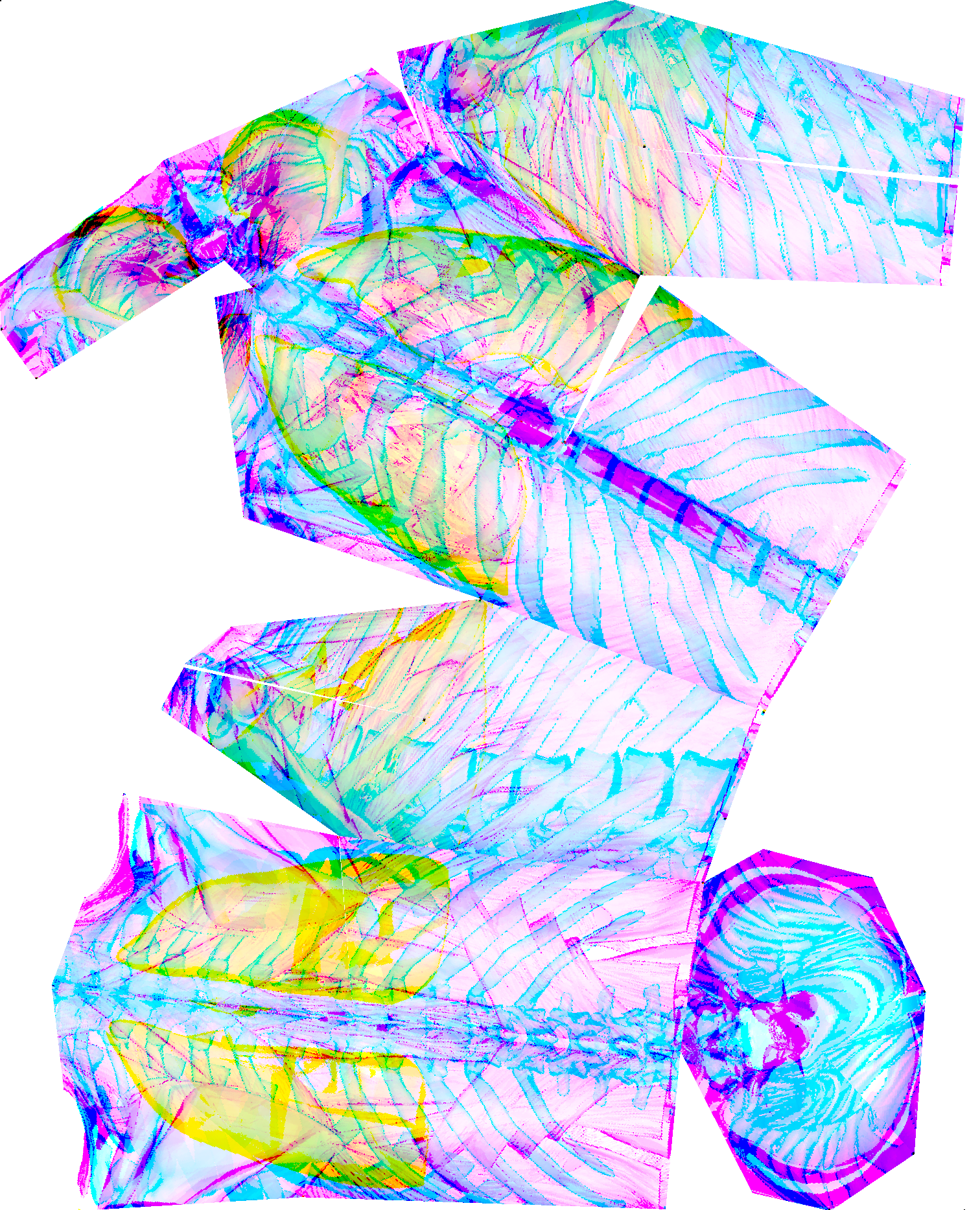} &
\includegraphics[height = 1.2in, width=0.12\textwidth]{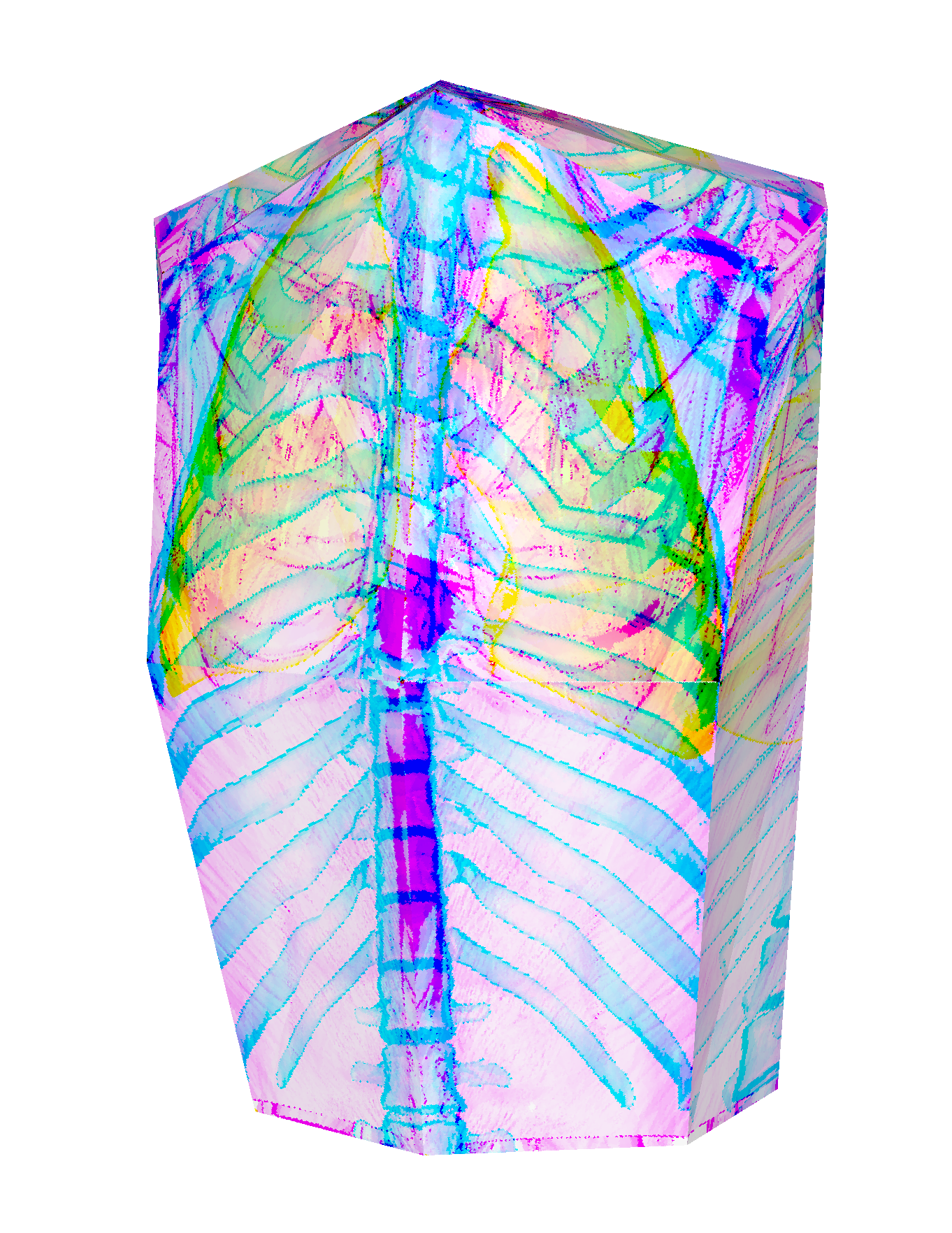} &
\includegraphics[height = 1.2in, width=0.10\textwidth]{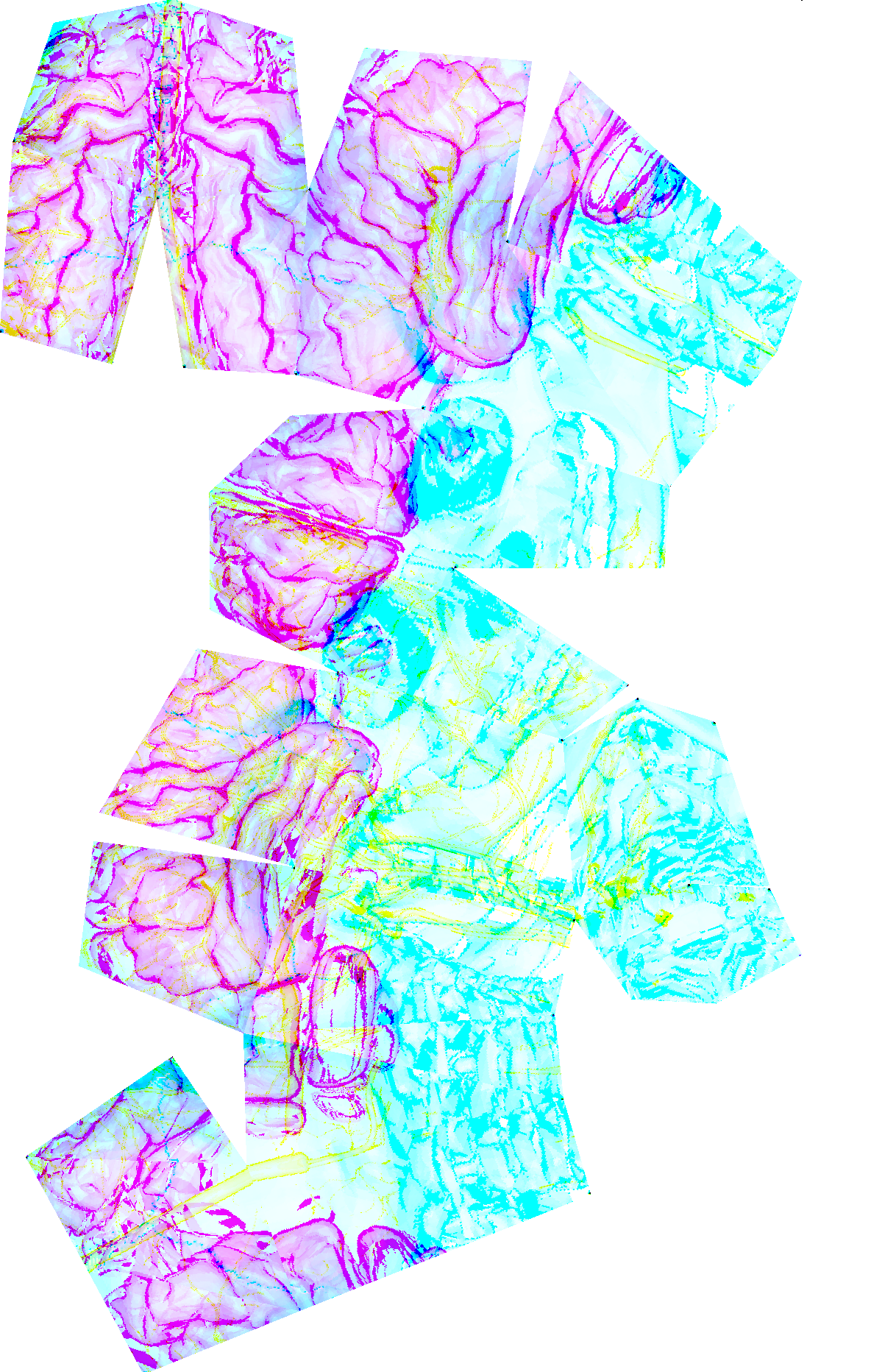} &
\includegraphics[height = 1.2in, width=0.13\textwidth]{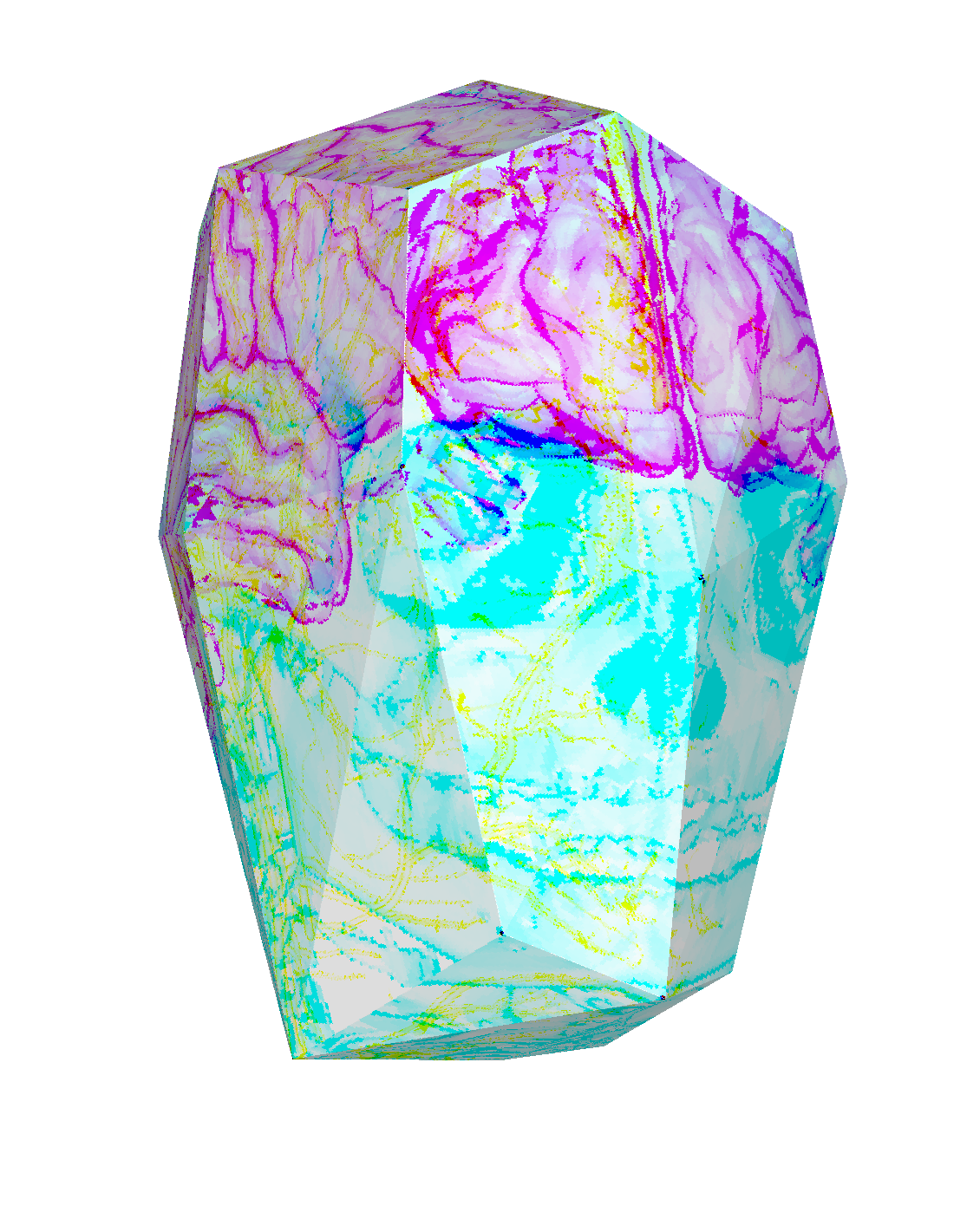} \\
\multicolumn{2}{c}{(a)} &
\multicolumn{2}{c}{(b)} \\
\end{tabular}
}
\caption{\vspace{-5pt}Textures and mesh unfoldings for (a) a torso and (b) a head.\vspace{20pt}}
\label{fig:result3db}
\centering{
\begin{tabular}{cc}
\includegraphics[height = 1.3in]{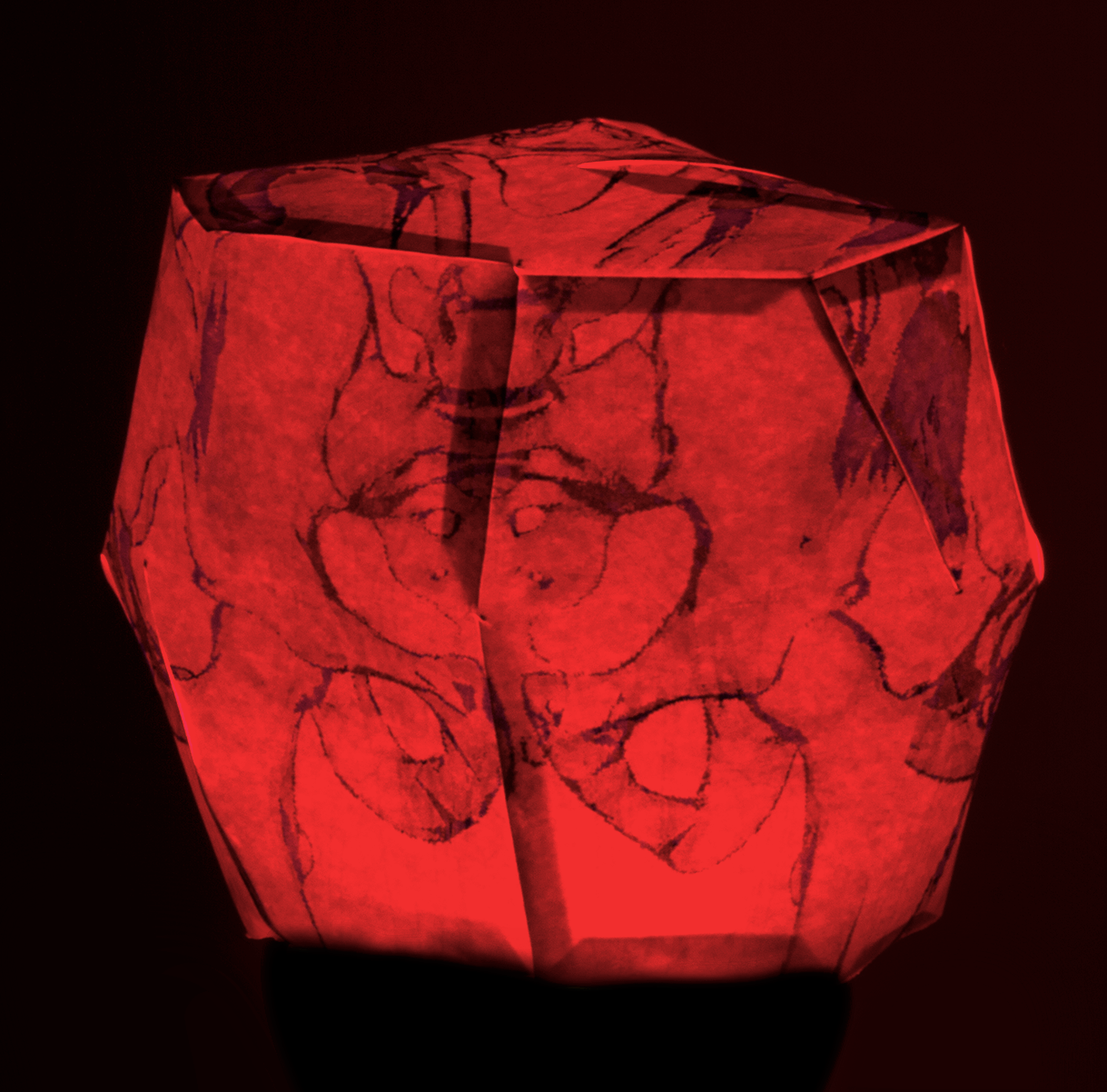} &
\includegraphics[height = 1.3in]{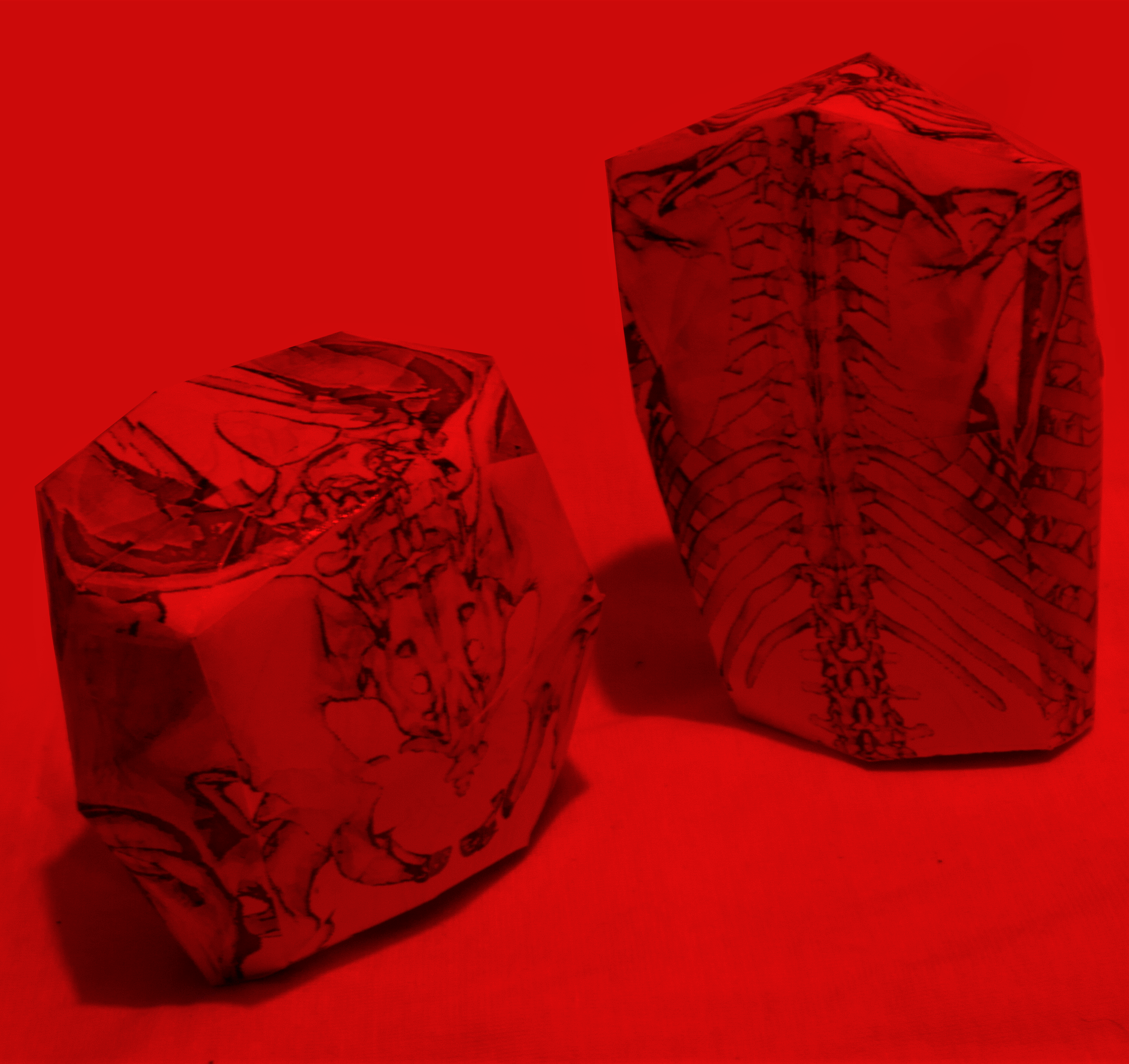}\\
(a) & (b)
\end{tabular}\vspace{-5pt}
}
\caption{Photos of 3D papercrafts with \XX{red} colored light placed (a) internally and (b) externally.\vspace{-20pt}}
\label{fig:light}
\end{figure}

\paragraph{Printable 2D Physical Visualizations:}
In Figure~\ref{fig:result2d}, we show several 2D printable physicalizations obtained for different parts of the human body (a-e). 
We see that the 2D case works well, \XX{regardless} of the underlying structures---as long as the color coding is appropriately chosen to match the available lenses.
The visibility of nested structures \XX{(}e.g., in the head or pelvis\XX{)} is further facilitated by the use of depth peeling, while brightening achieves the optimal hue with regard to color filters (Figure~\ref{fig:result2d}(f-g)). 
Color filters for isolating the underlying structures can also be previewed digitally (Figure~\ref{fig:result2d}(h)), as opposed to the physical filters shown in Figure~\ref{fig:teaser}(b).

\paragraph{Foldable 3D Physical Visualizations:}
In Figure~\ref{fig:result3da}, we show the results at each step of the workflow for the generation of a 3D foldable physicalization of a pelvis model. 
Here, the bones are encoded in cyan, the muscles in magenta and the vessels in yellow. 
As explained in Section~\ref{ssec:method}, a paper mesh template is generated, duplicated and wrapped around each structure (a). 
This is used to generate a combined texture (b), which is applied back to the (c) original paper mesh template for inspection. 
The respective unfolding can be printed and assembled into a 3D papercraft (d).
Two additional examples are shown in Figure~\ref{fig:result3db}, for the torso (a) and the head (b).
There are three interesting observations regarding the 3D papercrafts. 
First, due to the texture mapping through frame projection, there might be duplicates and artifacts in the final texture, which can be reduced with the help of the aforementioned local flattening.
Second, the appearance of the 3D papercrafts can be further improved to look less blocky, by using a larger number of triangles for the paper mesh. 
Sharp edges and strong changes in the normal vectors can be, thus, avoided. 
This, however, will incur additional computational efforts and time resources, to fold the papercraft~\cite{Korpitsch2020}. 
Third, additionally to colored lenses, colored light sources can be employed externally or internally, as seen in Figure~\ref{fig:light}, to bring forward individual structures. 
The lights could be used for larger models, or if the users need to keep their hands free.

\paragraph{Performance:} %
The workflow performs \XX{reasonably fast}, with the exception of the texture generation.
For example, for the torso dataset, the unfolding took $72$ seconds and the projection $26$ seconds, on a laptop with Intel Core i7 CPU (4 cores @2 GHz each) and 8 GB RAM. 
Refining the texture to eliminate perspective-related artifacts took approximately $5$ minutes, but this depends on how many changes need to be introduced.
With regard to the assembly of the 3D physical models (with $48$ triangles for the paper meshes in all examples), the torso papercraft took approximately $14$ minutes $30$ seconds for a pre-cut model, the head took $15$ minutes $20$ seconds and the pelvis took $16$ minutes $10$ seconds---indicating that time-wise the approach is reasonable. 
Yet, a thorough study needs to be conducted in the future, to assess the educational value and the overall feasibility of the approach. 
\section{Conclusions and Future Work} 
\label{sec:conclude}

We propose the \textit{Anatomical Edutainer} for the easy, accessible and affordable generation of 2D/3D physicalizations to be used in interactive and tangible anatomical edutainment. 
Our workflow supports the generation of printable 2D and foldable 3D physicalizations that change their visual properties and reveal different anatomical structures, under colored lenses or colored light.
Currently, the workflow has four limitations, which will be our directions for future work. 
First, we support the physical visualization of up to three structures, taking advantage of light color properties and of pre-determined colored filters. 
Investigating scalability to more than three structures and to additional filters is necessary. 
Second, in the 3D papercraft generation, automated minimization/correction of deformations and shape preservation of the paper mesh are anticipated to improve our current results, along with employing paper meshes with a larger number of triangles. 
Third, the unfolding step is currently performed in Blender and needs to be integrated into the application. 
\XX{Finally, further studies are required to evaluate the educational benefit and feasibility of the proposed physical visualization approach.}


\bibliographystyle{abbrv-doi}

\bibliography{paper}
\end{document}